\documentclass[11pt,twoside]{article}

\usepackage{asp2004}
\usepackage{epsf}
\usepackage{psfig}
\usepackage{lscape}

\newcommand{\xbootes}{XBo\"{o}tes}

\begin{document}
\title{X-ray and infrared properties of galaxies and AGNs in the 9
  deg$^2$ Bo\"{o}tes field}   
\author{R.C. Hickox$^1$, C. Jones$^1$, W.R. Forman$^1$, S.S. Murray$^1$, M. Brodwin$^2$,
  and the {\it Chandra} \xbootes, {\it Spitzer} IRAC Shallow Survey,
  AGES, and NOAO DWFS Teams}   
\affil{$^1$Harvard-Smithsonian Center for Astrophysics, $^2$Jet
  Propulsion Laboratory/Caltech}    

\begin{abstract} 
We examine the X-ray and infrared properties of galaxies and
AGNs in the 9 square degree Bo\"{o}tes field, using data from the {\it
  Chandra}
\xbootes\ and {\it Spitzer} IRAC Shallow Surveys, as well as optical
spectroscopy from the AGES survey.  A sample of $\sim$30,000 objects are
detected in all four IRAC bands, of which $\sim$2,000 are associated with
X-ray sources.  We also study X-ray fainter sources using stacking
techniques, and find that X-ray fluxes are highest for objects with
IRAC colors that are known to be characteristic of AGNs. Because these
are shallow, wide-field surveys, they probe the bright end of the AGN
luminosity function out to spectroscopic redshifts as high as $z=3$--4. We
can use this multiwavelength dataset to explore the properties and
redshift evolution of a large sample of luminous active galaxies.

\end{abstract}



\section{Overview}
Recent deep extragalactic surveys (e.g. GOODS, Lockman Hole, Groth
Strip) have provided great insights into distant galaxy populations,
in particular AGNs.  Deep surveys detect primarily faint, distant
objects, but miss many rare, more luminous objects.  A complementary
approach is to perform shallow, wide-field surveys that obtain large
numbers of sources and probe the bright end of the flux distribution.

The {\it Chandra} \xbootes\ (Murray et~al.\ 2005) and {\it Spitzer}
IRAC Shallow Surveys (Eisenhardt et~al.\ 2004) cover a large area of 9
deg$^2$ that also has deep optical photometry from the NOAO Deep Wide
Field Survey (Jannuzi \& Dey 1999), and optical spectroscopy from the
AGES survey (Kochanek et~al.\ 2004).  A total of $\sim$270,000 IRAC
sources are detected, of which $\sim$30,000 have detections in all
four IRAC bands.  Some $\sim$2000 of these have X-ray counterparts,
most of which are AGNs.  For many X-ray sources that do not have
optical spectroscopy, we also have photometric redshifts (Brodwin
et~al.\ 2006).  Here we provide a description of the datasets and
present some preliminary results.

\begin{figure}[t]
\centerline{\psfig{figure=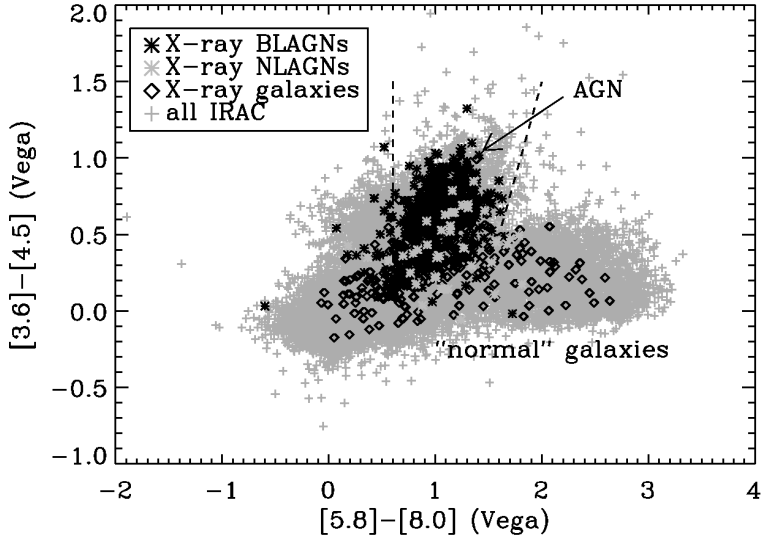,width=11cm}}
\caption{Color-color diagram for all IRAC sources detected at
  5$\sigma$ in
all four bands.  The Stern et~al.\ (2005) AGN region is shown by the
dashed lines.  X-ray sources with spectroscopic classifications
(broad-line AGNs, narrow-line AGNs, and normal galaxies (with
absorption line spectra)), are shown as stars.}  
\end{figure}

\section{Mid-IR selection of AGNs}
Using IRAC data and optical spectroscopy, Lacy et~al.\ (2004) and Stern
et~al.\ (2005) showed that AGNs can be selected on the basis of their
IRAC colors.  Fig.\ 1 shows the Stern et~al.\ (2005) color selection,
which was derived using data from Bo\"{o}tes.  The IR colors of
\xbootes\ X-ray sources with optical spectroscopy are shown.  Optical
spectra are classified as broad-line AGNs, narrow-line AGNs, or
``normal'' galaxies.  X-ray sources, particularly those classified as
broad-line AGNs, lie preferentially in this region of color-color
space.

\begin{figure}[t]
\hbox{
\psfig{figure=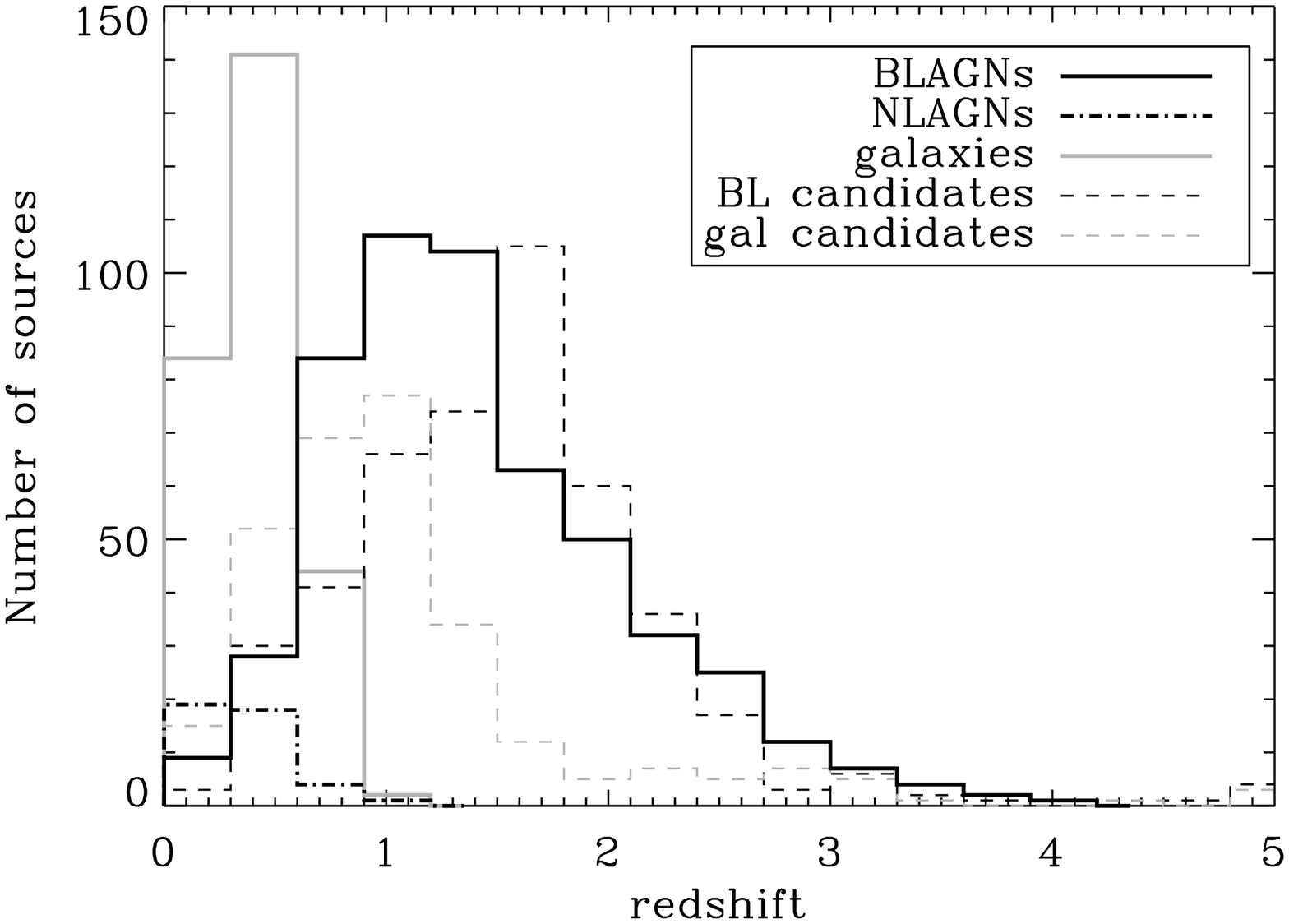,width=6.2cm}
\hspace{0.0cm}
\psfig{figure=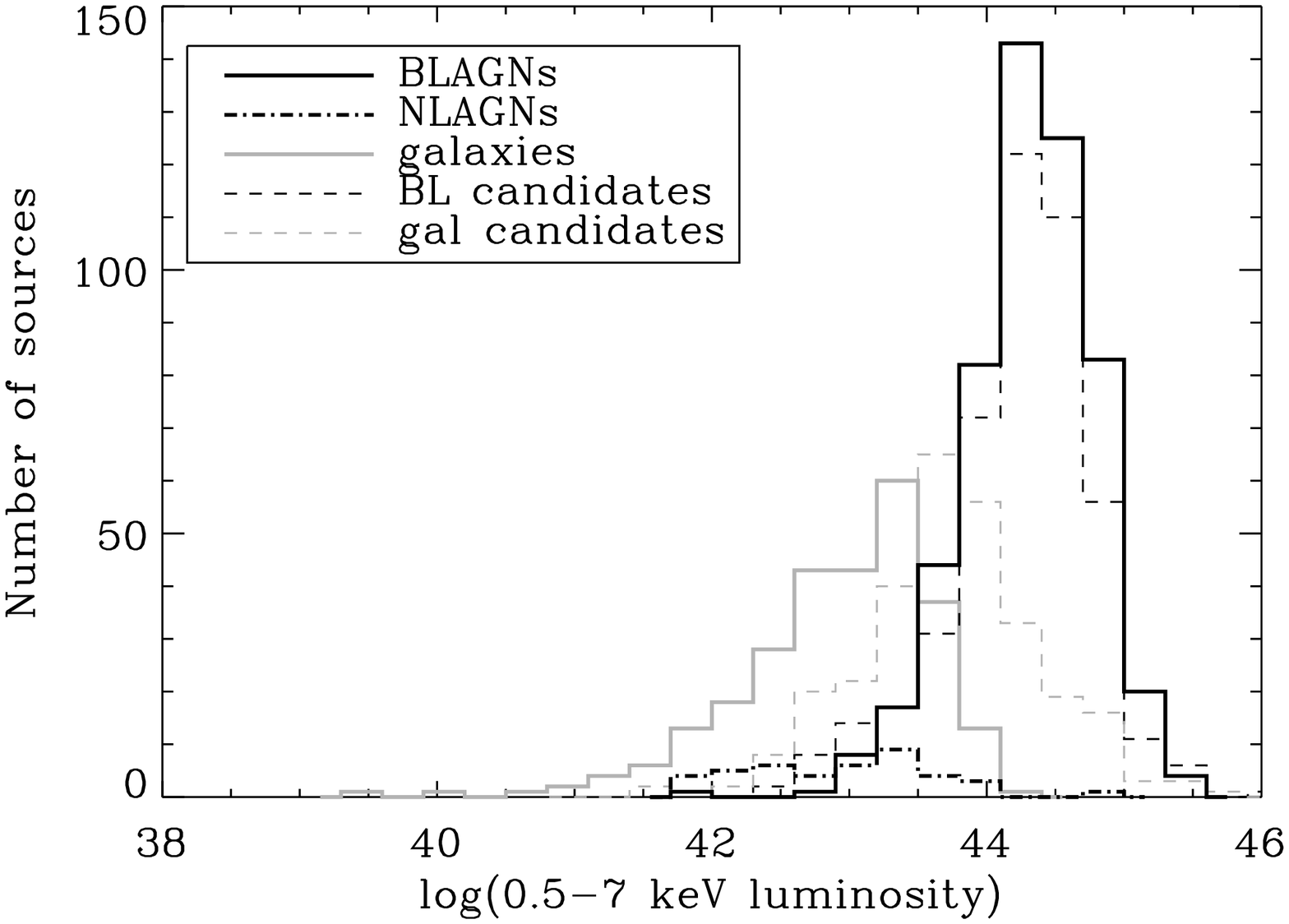,width=6.2cm}
}
\caption{Left: redshift distribution for all the X-ray sources in the
  \xbootes\ catalog with spectroscopic or photometric redshifts.  The
  different optical spectroscopic classifications are shown as bold lines, while objects with photo-z's are classified as AGN or normal
  galaxies by the Stern et~al.\ (2005) color selection (see Fig.\ 1),
  and are shown with light dashed lines.  Right: 0.5--7 keV X-ray luminosity
  distribution for X-ray sources with redshifts.  A $\Gamma=1.7$ power
  law with Galactic absorption is assumed. }
\end{figure}

\section{Redshift and luminosity distributions}
Fig.\ 2 shows the redshift and $L_{\rm X}$ distributions for all the
X-ray sources in the \xbootes\ catalog with redshifts.  Note that
``normal'' galaxies are not seen above $z \sim 1$, however AGNs are
detected out to $z \sim3$--4.  Because of the wide field and flux
limits of the \xbootes\ survey, we detect preferentially distant,
luminous AGNs with $L_{\rm X} > 10^{44}$ ergs.

\begin{figure}[t]
\hbox{
\psfig{figure=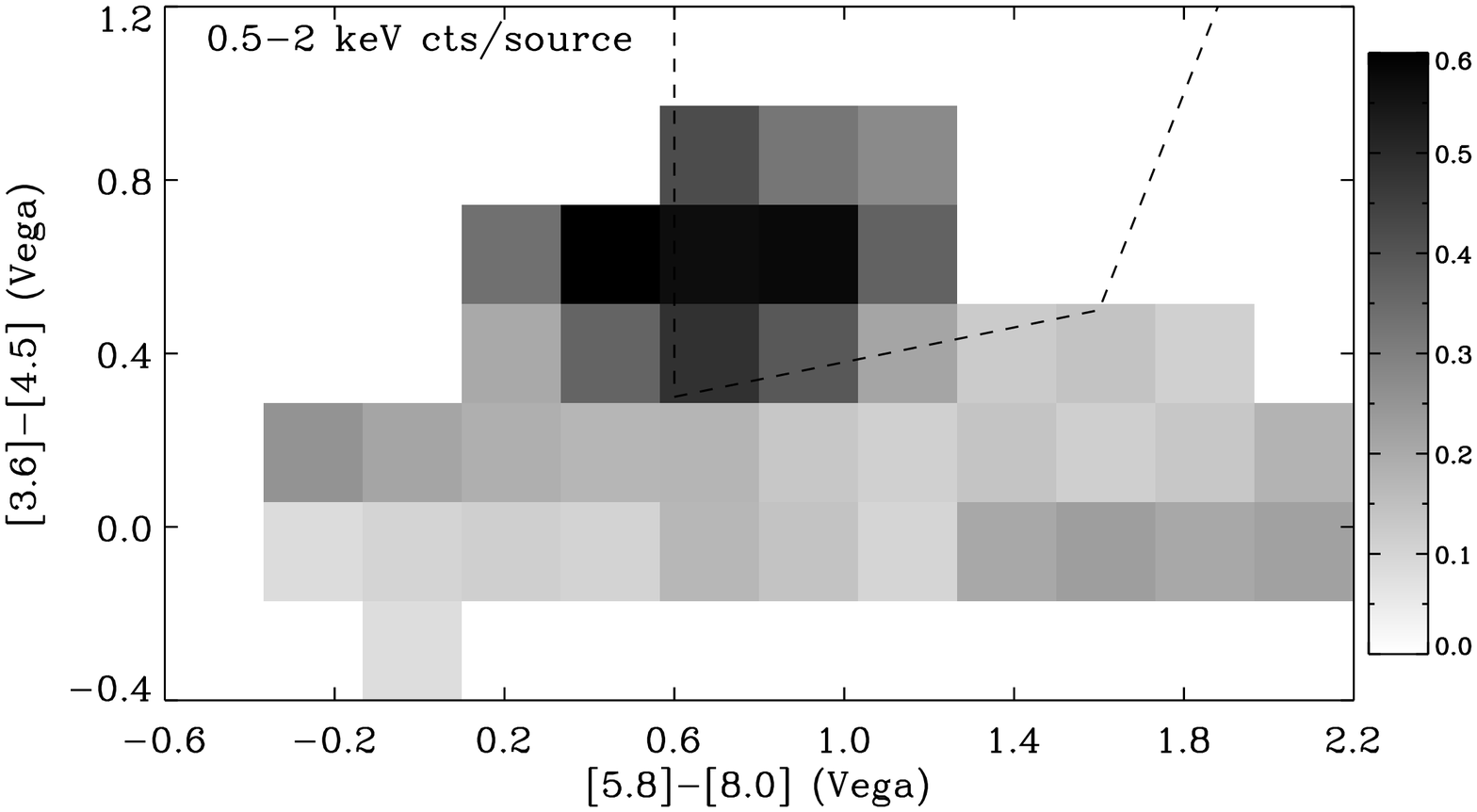,width=5.9cm}
\hspace{0.5cm}
\psfig{figure=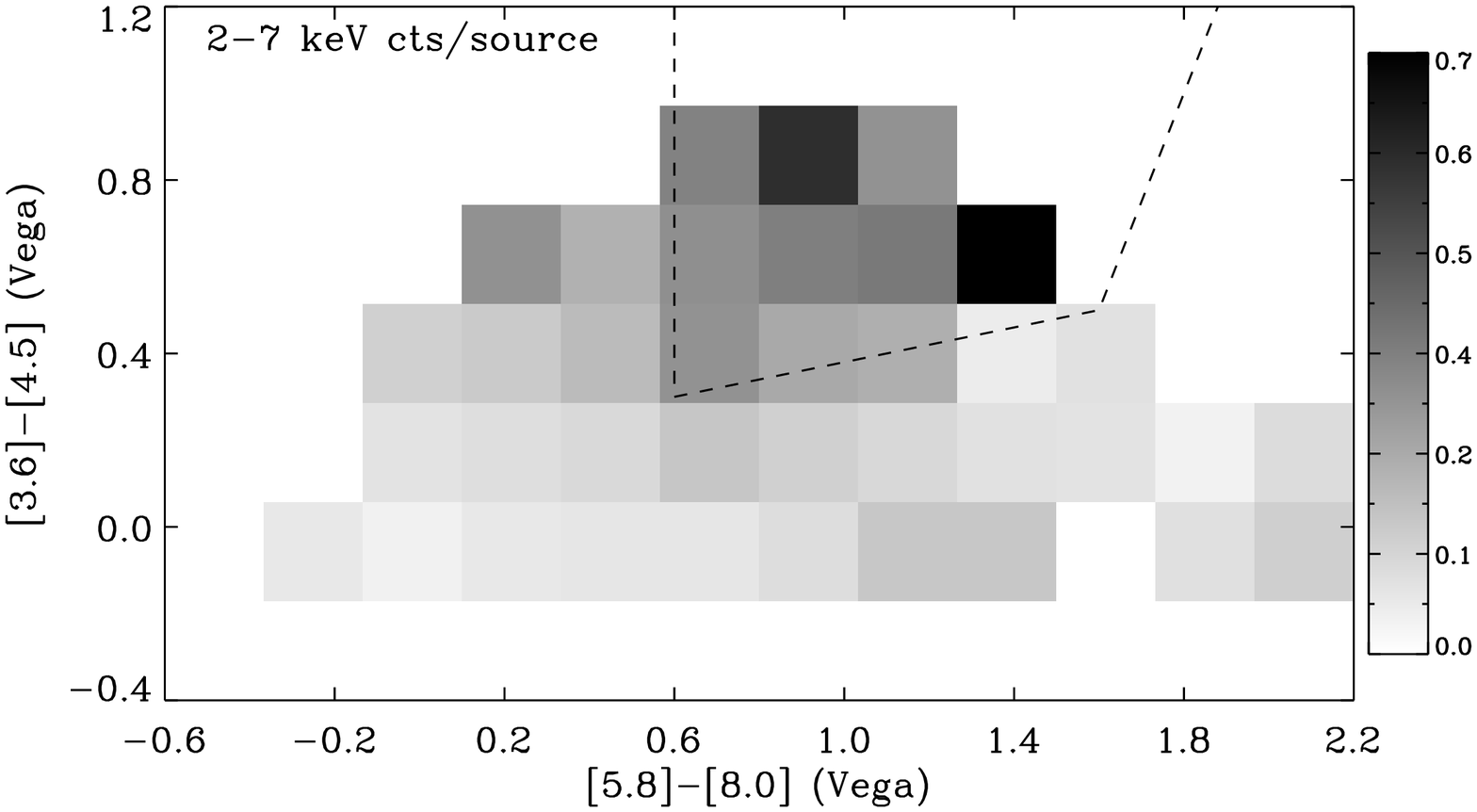,width=5.9cm}
}
\caption{Average X-ray emission (in counts/source) for IRAC sources
  that do not have X-ray matches,  binned by their IR colors for (left)
  0.5--2 keV and (right) 2--7 keV.  The Stern et~al.\ (2005) AGN region
  is shown as in Fig.\ 1.}
\end{figure}

\section{X-ray stacking analysis}
For objects that do not have detected X-ray counterparts, we can study
their average X-ray emission by stacking analyses.  Fig.\ 3 shows the
average X-ray flux as a function of IR color.  Note that in both soft
and hard X-rays, the average photon flux is $<$1 count per source.
However, the emission peaks in the AGN region defined by Stern
et~al.\ (2005), indicating a population of X-ray emitting AGNs at below
the \xbootes\ flux limits.

\section{Prospects}
The Bo\"{o}tes sample of thousands of X-ray, infrared, and optically
selected active galaxies will allow us to address a number of
questions such as:
\begin{enumerate}
\item How do the broad-band spectral properties and number densities
of luminous AGNs evolve with redshift? Are these related to changes in
the accretion process, particularly for $z < 1$ where luminous AGN
activity may decline sharply (e.g. Barger 2005)?

\item Are significant numbers of AGNs obscured in the optical or X-ray
bands?  What multiwavelength properties of these sources can be used
to identify them?

\item How are the properties of active galaxies related to their local
environment?  For the Bo\"{o}tes field, local galaxy densities can be
determined by the 3-D distributions from the AGES data, and correlated
to the luminosities and spectral types of AGNs.  Do such correlations
change with redshift?

\end{enumerate}



\end{document}